\documentclass[twocolumn]{article} 

\usepackage{preprint}

\usepackage{amsmath, amsthm, amssymb, amsfonts}
\usepackage[numbers,square]{natbib}

\usepackage[utf8]{inputenc}	
\usepackage[T1]{fontenc}	
\usepackage{xcolor}		
\usepackage[colorlinks = true,
            linkcolor = purple,
            urlcolor  = blue,
            citecolor = cyan,
            anchorcolor = black]{hyperref}	
\usepackage{booktabs} 		
\usepackage{nicefrac}		
\usepackage{microtype}		
\usepackage{lineno}		
\usepackage{float}			

\usepackage{lipsum}		

\usepackage{newfloat}
\DeclareFloatingEnvironment[name={Supplementary Figure}]{suppfigure}
\usepackage{sidecap}
\sidecaptionvpos{figure}{c}

\usepackage{graphicx}
\usepackage{natbib}
\usepackage{doi}
\usepackage{algorithm}
\usepackage{textcomp, gensymb}

\usepackage[noend]{algpseudocode}

\usepackage{xcolor}
\usepackage{graphicx}
\usepackage{lineno}
\usepackage{multirow}
\usepackage{stfloats}
\usepackage[normalem]{ulem}
\usepackage{xurl}
\usepackage{pdfpages}

\newcommand{\modulus}{ \ensuremath{\D a}}

\newcommand{\phase}{ \ensuremath{\phi}}
\newcommand{\transmittance}{ \ensuremath{ \U{\D{t}}}}
\newcommand{\hologram}{\ensuremath{\D{d}}}

\newcommand{\propagator}{\ensuremath{\D{\U{h}}}}
\newcommand{\SVPSFpropagator}{\ensuremath{\V{\D{\U{h}}}}^\textmd{SV}}
\newcommand{\SVPSFpropagatornb}{\ensuremath{{\D{\U{h}}}}^\textmd{SV}}
\newcommand{\SIPSFpropagator}{\ensuremath{\V{\D{\U{h}}}_z^\textmd{SI}}}

\newcommand{\PSF}{\ensuremath{\U{\D{p}}}_0}

\newcommand{\Udiff}{\U{{\D{U}}}_{\textmd{diff}}}

\newcommand{\V}[1]{\ensuremath{\boldsymbol{#1}}} 
\newcommand{\U}[1]{\ensuremath{\underline{#1}}} 
\newcommand{\D}[1]{\ensuremath{#1}} 

\newcommand{\rev}[1]{\textcolor{black}{#1}}

\usepackage{titlesec}
\titlespacing\section{0pt}{12pt plus 3pt minus 3pt}{1pt plus 1pt minus 1pt}
\titlespacing\subsection{0pt}{10pt plus 3pt minus 3pt}{1pt plus 1pt minus 1pt}
\titlespacing\subsubsection{0pt}{8pt plus 3pt minus 3pt}{1pt plus 1pt minus 1pt}

\usepackage{tikz,xcolor,hyperref}

\definecolor{lime}{HTML}{A6CE39}
\DeclareRobustCommand{\orcidicon}{
	\begin{tikzpicture}
	\draw[lime, fill=lime] (0,0) 
	circle [radius=0.16] 
	node[white] {{\fontfamily{qag}\selectfont \tiny ID}};
	\draw[white, fill=white] (-0.0625,0.095) 
	circle [radius=0.007];
	\end{tikzpicture}
	\hspace{-2mm}
}
\foreach \x in {A, ..., Z}{\expandafter\xdef\csname orcid\x\endcsname{\noexpand\href{https://orcid.org/\csname orcidauthor\x\endcsname}
			{\noexpand\orcidicon}}
}

\title{Iterative phase retrieval algorithm for space-variant PSF in optical systems with aberrations}

\usepackage{authblk}

\author[1,*]{Dylan Brault\orcidA{}}
\author[1]{Corinne Fournier\orcidB}
\author[2,3]{Tatiana Latychevskaia\orcidC}

\affil[1]{Université Jean Monnet Saint-Etienne, CNRS, Institut d'Optique Graduate School, Laboratoire Hubert Curien UMR 5516, 42023, Saint-Etienne, France}
\affil[2]{Center for Life Sciences, Paul Scherrer Institute, Forschungsstrasse 111, 5232 Villigen, Switzerland.}
\affil[3]{Department of Physics, University of Zurich, Winterthurerstrasse 190, 8057 Zurich, Switzerland}
\affil[*]{dylan.brault@telecom-st-etienne.fr}

\begin{document}

\twocolumn[ 
  \begin{@twocolumnfalse} 
  
\maketitle

\begin{abstract}
Iterative phase retrieval algorithms are widely used in digital optics \rev{for their efficiency and simplicity}. Conventionally, these algorithms do not consider aberrations as they assume an ideal, aberration-free optical system. Here, we propose modified iterative phase retrieval algorithms that take into account the space-invariant and space-variant point spread function of the optical system.
\end{abstract}
\vspace{0.35cm}

  \end{@twocolumnfalse} 
] 



\section{Introduction}
Iterative Phase Retrieval (IPR) algorithms \cite{fienup1978reconstruction,gerchberg1972practical} allow reconstruction of the complete complex-valued wavefront from one or more intensity measurements by using back-and-forth wavefront propagation between the detector and sample's planes and applying constraints $\gamma$ based on the object's properties (most commonly, finite size and positive absorption). Originally, IPR algorithms were applied for the reconstruction of objects from X-ray diffraction patterns, for no lenses are available for X-rays \cite{miao1999extending}. Later, the IPR approach was generalized to digital in-line holography, where it allowed to solve the twin image problem \cite{latychevskaia2007solution}. In the case of in-line holography, the wavefront propagation is often calculated through a convolution with a free-space propagation function. A hologram can be obtained using either a lens-free system, as in the case of in-line, Gabor type or low-energy electron holography \cite{latychevskaia2019direct,longchamp2017imaging}, or by using an optical system with lenses, as in the case of digital holographic microscopy imaging \cite{kim2010principles}. Since there are no perfect lenses, such optical systems always have aberrations. In this study, we show how an iterative phase retrieval algorithm can be modified by including aberration correction directly into the algorithm. The proposed method allows considering both space-invariant and space-variant Point Spread Functions (PSF). We describe the principle and the algorithms, and as an example of practical application, we demonstrate the results obtained by applying a modified IPR algorithm for light optical holograms. The proposed here approach can be applied for optical systems with aberrations and/or space-variant PSF. It could be potentially useful for application in electron microscopy, where the resolution is mainly limited by the lens aberrations.
\section{Hologram formation models}\label{sec:HoloModels}
In this section, the conventional model of wavefront propagation through an ideal optical system is first recalled. Then, a wavefront propagation model that accounts for space-invariant aberration is introduced. Finally, we present a method to propagate a wavefront numerically with space-variant aberrations at a low computational cost. 
\subsection{Hologram formation model without aberration} \label{sec:Withoutaberrations}
Reconstruction algorithms aim to retrieve the complex-valued transmission function that describes how the incident wave and the sample interact. The transmission function is typically modeled as a 2D discrete complex-valued distribution, with its modulus, $\V{\modulus}$, describing the absorption properties of the sample and phase $\V{\phase}$ describing the phase delay added by the sample to the incident wavefront:
\begin{equation}
    {\transmittance}(x,y)={\modulus}(x,y) \exp(i {\phase}(x,y))
\end{equation}
 where $(\D x,\D y)$ are the coordinates in the sample plane and the underlyed values are complex-valued distributions.
When a plane wave of unitary amplitude illuminates the sample, in the absence of noise, the data $\V \hologram$ recorded on the sensor plane, the so-called hologram, corresponds to the intensity of the interference pattern of the sample at a distance $z$ and is modeled by:
\begin{equation}
    {\hologram}({\D x', \D y'})=\lvert \Udiff({\D x', \D y'})  \rvert^2=\lvert \V{\D{\transmittance}} \underset{\D x', \D y'}{\ast} \V{\D{\propagator}_z}   \rvert^2
\end{equation}
where $\Udiff$ is the discrete complex amplitude of the diffracted wavefront, $\V{\D{\transmittance}}$ is the transmission function, $\V{\D{\propagator}}_z$ is a discrete propagation kernel for a distance $z$, $(\D x',\D y')$ are the coordinates in the hologram plane and $\ast$ is the discrete convolution operator. 
For light optical holograms, Fresnel and Rayleigh-Sommerfeld kernels are commonly used to simulate the wavefront propagation \cite{goodman2005introduction}. 
This hologram formation model describes the propagation in a perfect optical system and is the core of most reconstruction algorithms. However, slight misalignment, tilts of the sample, nonstandard uses of microscope objectives, or poorly corrected objectives lead to aberrations that must be considered to perform morphologically reliable and quantitative sample reconstructions. 

\subsection{Model of propagation with a Space-Invariant PSF (SI-PSF)} \label{sec:SI-PSFmodel}
For PSF that do not vary in the field of view (FOV), wavefront propagation can still be modeled as a convolution between the transmission function $\transmittance$ and a new aberrated propagation kernel $\SIPSFpropagator$. This aberrated propagation kernel is given by the convolution between the aberration-free propagation kernel and the complex-valued PSF of the optical system $\V{\PSF}$ for $z=0$:
\begin{equation}
    \SIPSFpropagator(\D{x}',\D{y}')=\V{\D{\propagator}}_z \underset{\D{x}',\D{y}'}{\ast} \V{\PSF}
\end{equation}
$\V{\PSF}$ for $z=0$ can, for example, be estimated with model fitting approaches like proposed in \cite{brault2022accurate}.
Computing the propagation of a wavefront with a space-invariant aberration has the same computational cost as conventional aberration-free propagation, because the wavefront propagation model remains a simple convolution.

\subsection{Model of propagation with a Space-Variant PSF (SV-PSF)} \label{sec:SV-PSFmodel}
For PSF that vary in the FOV, one has to interpolate the complex-valued PSF to compute the image formation model. Naive interpolation of the PSF is prohibitive for computational time reasons as it would require a convolution with a local propagation kernel for each image pixel. Several techniques have been developed to reduce the computational time of PSF interpolation \cite{nagy1998restoring,flicker2005anisoplanatic,hirsch2010efficient,denis2015fast}. Most of these techniques have been applied to compute space-variant aberration models with real-valued PSF and real-valued objects. Also, contrary to holography applications, which require extended propagation kernels, these methods have been mostly used for PSF located on a small spatial support. We chose to adapt the method proposed by Flicker and Rigaut \cite{flicker2005anisoplanatic} to interpolate the PSF. \\
This method relies on a modal decomposition of the space-varying PSF $\V{\D{\SVPSFpropagator}}$:
\begin{equation}
    \D{\SVPSFpropagatornb}(\D x,\D y,\D x',\D y') \approx \sum_{p=1}^{P} \U{\D{w}}_p(\D{x},\D{y})\U{\D{m}}_p(\D{x}',\D{y}')
    \label{eq:SeparableApprox}
\end{equation}
where $P$ is the number of spatial locations where the PSF has been estimated, $\U{\V{\D{m}}}=\{\U{\V{\D{m}}}_p\}_{p=1,...,P}$ is a set of space-invariant orthogonal components of the PSF (modes), and $\U{\V{\D{w}}}=\{\U{\V{\D{w}}}_p\}_{p=1,...,P}$ is a complex-valued weighting map that describes how the mode $\U{\D{m}}_p$ should be weighted in the FOV. This separable approximation simplifies the image formation model such that the complex amplitude in the detector plane $\Udiff$ can be expressed as a sum of convolutions:
\begin{equation}
    \Udiff(\D{x}',\D{y}')=\SVPSFpropagator \underset{\underset{(\D{x}',\D{y}')}{\textmd{SV}}}{\ast} \transmittance =\sum_{p=1}^{P} \left[ \V{\U{\D{w}}}_p \odot \V{\D{\transmittance}} \right]  \underset{\D x', \D y'}{\ast}  \V{\U{\D{m}}}_p
    \label{eq:SVtransmittance}
\end{equation}
where $\underset{\textmd{SV}}{\ast}$ symbolizes the space-variant convolution and $\odot$ is the pixel-wise product.\\
In practice, only a few modes are required when \D{\SVPSFpropagator} slowly varies in the FOV. Thus, the separable approximation of  Eq. \ref{eq:SeparableApprox} can be performed using a low-rank approximation (see Fig. S2) \rev{as proposed in \cite{flicker2005anisoplanatic}}. The matrix $\U{\V{K}}$, whose columns contain each estimated PSF, is decomposed using an Singular Value Decomposition (SVD):
\begin{equation}
    \U{\V{K}}=\U{\V{U}}\cdot \V{\Sigma} \cdot \U{\V{V}}^\top=\sum_{p}^{P} \U{\V{{u}}}_p\cdot \V{\sigma}_p \U{\V{v}}_p^\top\approx \sum_{p=1}^{Q} \U{\V{{u}}}_p \cdot \V{\sigma}_p \U{\V{v}}_p^\top
    \label{eq:LowRankDecomp}
\end{equation}
where $\U{\V{{u}}_p}$ and $\U{\V{v}_p}$ are the $p$-th left and right singular vectors, $\V{\sigma}_p$ the corresponding singular value \rev{and $\cdot$ is the matrix multiplication}. By analogy with Eq. \ref{eq:SeparableApprox}, $\U{\V{{U}}}$ is a matrix that contains the orthogonal modes $\U{\V{\D{m}}}$ on which the PSF are decomposed and $\V{\Sigma} \cdot \U{\V{V}}^\top$ contains the corresponding weights $\U{\V{\D{w}}}$. As the singular value $\V{\Sigma}$ decreases quickly, a low-rank approximation can be performed with $Q < P$. Moreover, considering the complex-valued PSF is slowly varying in the FOV, instead of estimating the PSF on each pixel, only an estimation on a coarse regular grid is required. Then, a smooth interpolation of weights $\U{\V{\D{w}}}$ on the pixel grid is performed such that each $\U{\V{\D{w}}}_p$ has the same size as $\transmittance$. We chose to interpolate these weights with cubic functions $\varphi$. 
Note that in the following sections, the back-propagation kernel associated with $\V{\propagator}$ is noted $\V{\propagator}^\dagger$.
\section{Generalization of the IPR algorithm for aberrated systems} \label{sec:PhaseRetrieval_Alg}

In the following, we describe how to adapt the IPR algorithm for optical systems with SI-PSF and SV-PSF.
Alg. S1 reminds the pseudo code of the IPR algorithm~\cite{latychevskaia2007solution}, and Fig. S1 depicts a general scheme of this approach.
\subsection{Iterative phase retrieval algorithm for SI-PSF} \label{sec:PhaseRetrieval_SI-PSF}
In the case of SI-PSF, the IPR algorithm is modified by replacing the aberration-free propagator with $\SIPSFpropagator$. The backpropagation step is performed by convolving the complex amplitude $\Udiff$ with ${\SIPSFpropagator}^{\dagger}$ where ${\SIPSFpropagator}^{\dagger}$ is obtained by conjugating ${\SIPSFpropagator}$ in the Fourier domain such that convolving $\SIPSFpropagator$ with ${\SIPSFpropagator}^{\dagger}$ is approximately an identity operation (sampling and truncation of the FOV aside).
The IPR algorithm in this configuration is very similar to the classical IPR algorithm \cite{latychevskaia2007solution}. Thus, after calibrating the PSF, the computational time needed for the reconstruction is the same as that required for the classical IPR algorithm.
 
\subsection{Iterative phase retrieval algorithm for SV-PSF} \label{sec:ERSV-PSF}
To account for SV-PSF in the IPR algorithm, we propose to exploit the image formation model described in Sec. \ref{sec:HoloModels}.\ref{sec:SV-PSFmodel} to perform the propagation steps. The backpropagation step is performed by conjugating the modes $\U{\V{\D{m}}}$ in the Fourier domain and by conjugating the weights $\U{\V{\D{w}}}$. Alg. \ref{alg:ERSV-PSF} provides a pseudo code for the IPR algorithm considering SV-PSF. Following Eq. \ref{eq:SVtransmittance}, the propagation steps are performed as the sum of weighted convolution instead of a simple convolution. The computational time of the IPR algorithm is then proportional to the rank $Q$ of the low-rank approximation described in Eq. \ref{eq:LowRankDecomp}.

\begin{algorithm}

\caption{IPR algorithm for SV-PSF}\label{alg:ERSV-PSF}
\begin{algorithmic}[1]
\State \textbf{Input:} Data $\V{\hologram}$ \hfill ($L \times C$ image)
\State \textbf{Input:} Propagation kernels matrix $\U{\V{K}}$ \hfill ($LC  \times P$ matrix)
\State \textbf{Input:} Number of iterations $N_{\textmd{iter}}$ \hfill (integer)
\State \textbf{Input:} Rank of the low rank approximation $Q$ \hfill (integer)
\State \textbf{Output:} Transmission function $\V{\transmittance} \hfill $ ($L \times C$ image)
\Statex
\Statex \makebox[\linewidth]{ \hfill \underline{Computation of modes and weights}}
\Statex
\State $[\U{\V{U}},\U{\V{\Sigma}},\U{\V{V}}] \gets \textmd{SVD}(\U{\V{K}})$ \Comment{Low rank approximation of $\U{\V{K}}$}
\State $\U{\V{\D{w}}} \gets \varphi(\U{\V{\Sigma}} \cdot \U{\V{V}}^\top)$ \Comment{Interpolating weights on the sensor grid} 
\State $\U{\V{\D{m}}} \gets \texttt{reshape}(\U{\V{U}})$ \Comment{Reshaping $\U{\V{U}}$($\rightarrow L \times C\times P$ 3D matrix)} 
\Statex
\Statex \makebox[\linewidth]{ \hfill \underline{IPR algorithm for SI-PSF}}
\Statex
\State $\V{\Udiff} \gets \sqrt{\V{\hologram}}$ \Comment{Initialization of the wavefront on the sensor} \\
\For {$n=1 \textbf{ to } N_{\textmd{iter}}$}
\State $\V{\transmittance} \gets \sum_{p=1}^{Q} (\U{\V{\D{w}}}_p^{\ast} \odot \V{\Udiff}) \ast \U{\V{\D{m}}}_p^{\dagger}  $  \Comment{Back-propagation of $\Udiff$} \\
\For{$k=1 \textbf{ to } L \times C$}
\If{$\transmittance_k \notin \gamma$}  \Comment{Applying constraints} 
\State $\transmittance_k=1$    
\EndIf
\EndFor\\
\State $\V{\Udiff} \gets  \sum_{p=1}^{Q} (\U{\V{\D{w}}}_p \odot \V{\transmittance}) \ast\U{\V{\D{m}}}_p  $ \Comment{Propagation of $\transmittance$} \\
\State $\V{\Udiff} \gets \sqrt{\V{\hologram}} \times \exp(i \times \texttt{arg}(\V{\Udiff}))$ \Comment{Updating modulus} \\
\EndFor
\State \textbf{return} $\V{\transmittance}$
\end{algorithmic}
\end{algorithm}

\section{Results} \label{sec:Results}
The proposed approach has been applied to simulated and experimental light optical holograms. In the results presented below, the propagation was simulated using Rayleigh-Sommerfeld approximation \cite{goodman2005introduction}.
\subsection{Reconstruction results on simulated holograms} \label{sec:SimuResults}
To test the robustness of our approach, two holograms where simulated  with the model for SV-PSF described in section~\ref{sec:HoloModels}.\ref{sec:SV-PSFmodel}. The first hologram was simulated with a slowly spatially varying PSF and the second one with a rapidly varying PSF. These two SV-PSF holograms are reconstructed under three described model hypothesis. 

\begin{figure}[htbp]
    \centering
    \includegraphics[width=0.82\linewidth]{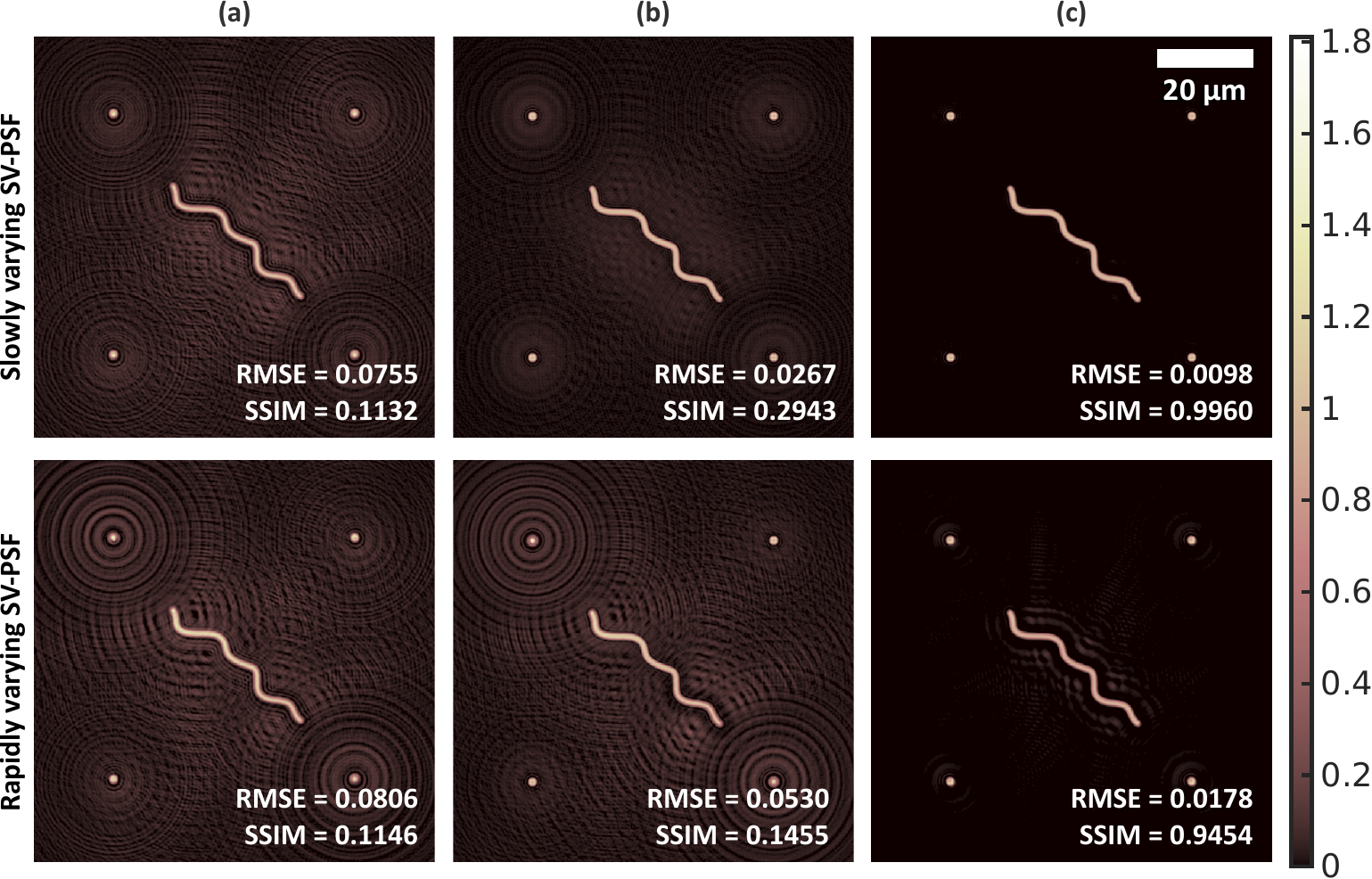}
    \caption{Phase distributions reconstructed from the simulated hologram (a) with classical IPR algorithm, (b) with SI-PSF IPR algorithm, and (c) with SV-PSF IPR algorithm ($P=49,Q=15$)}
    \label{fig:Simu_Recons}
\end{figure}

A pure phase binary object with maximal phase shift of 1 radian was used in the simulations (see Fig.\ref{fig:Simu_Recons}). To perform realistic hologram simulations, the experimental PSF of the hologram described in section \ref{sec:Results}.\ref{sec:ExpeResults} was used. 
The PSF \textcolor{black}{(defocus and aberrations)} was estimated at $P=49$ positions ($7 \times 7$ regular grid) in the FOV using the method described in \cite{brault2022accurate}. \textcolor{black}{Typical Zernike coefficients are provided in Supplementary Information (Table S2).} As this method estimates the PSF on an irregular grid, interpolating the estimated Zernike coefficients on the regular grid has been performed using LOWESS interpolation \cite{cleveland1979robust}. The simulation parameters (matching the experimental ones) are provided in Table S1 and the pupil functions are displayed in Fig. S3. 
In this case, aberrations are slowly varying, furthermore, the sample slide is slightly tilted (0.5\degree \ tilt). 
Simulations with these experimental PSFs are referred to as 'Slowly varying SV-PSF' simulations in the following.
To study the efficiency of our approach, we performed reconstruction with a more rapidly varying SV-PSF that was achieved by tilting the sample slide of 3.5\degree. \textcolor{black}{Even though this tilt introduces a linear spatial variation in $z$, the PSF, which is a Rayleigh-Sommerfeld kernel, does not evolve linearly.}
Simulations with these PSFs are referred to as 'Rapidly varying SV-PSF' simulations in the following.
The classical IPR algorithm reconstructions have been performed using a distance measured with a sharpness autofocusing criterion \cite{zhang2017edge}, so that the evaluated reconstruction errors do not come from a defocus error due to spherical aberrations.
As the phase object introduces a positive phase shift, the sample constraint $\gamma$ has been chosen such that the reconstructed object should be a pure phase with a positive phase shift. To achieve convergence of all the proposed reconstruction methods, we set the number of iterations $N_\textmd{iter}=200$. 
Fig. \ref{fig:Simu_Recons}.(a) shows the reconstructions of the simulated SV-PSF holograms with the classical IPR algorithm and accounting for SI-PSF and SV-PSF presented in section~\ref{sec:PhaseRetrieval_Alg}. To provide a quantitative comparison the root mean square error (RMSE) and the structural similarity index measure (SSIM) were estimated between the reconstructed distributions and the ground truth.
Fig. \ref{fig:Simu_Recons}.(a) shows that the classical IPR algorithm fails to reduce the twin image in both simulated cases, leading to high error values and strong reconstruction artifacts. Compared to the other methods, the classical IPR algorithm provides the worst RMSE and SSIM.
As expected, considering an SI-PSF (Fig. \ref{fig:Simu_Recons}.(b)) with aberrations estimated at the center of the FOV improves the reconstruction quality for slowly varying PSF. However, when the PSF variation increases, the SI-PSF hypothesis is insufficient to reduce the twin image artifacts.
As expected, considering the SV-PSF (Fig. \ref{fig:Simu_Recons}.(c)), the quality of the reconstructions regarding RMSE and SSIM has increased significantly. The twin image artifacts are almost entirely removed in that case, leading to more accurate reconstructions. Only $Q=15$ modes have been used to model the SV-PSF in these reconstructions. However, let us notice that the reconstruction quality could be further improved by increasing $Q$. \\
As computational time is the bottleneck of our SV-PSF method, we proposed reconstructing the slowly varying SV-PSF simulated hologram with a limited convolution budget. Thus, we chose $n_\textmd{iter}$ and $Q$ such that the algorithm should stop after 400 convolutions. We compared the reconstruction results with the classical IPR algorithm and SI-PSF IPR algorithm with 400 convolutions (see Table \ref{tab:budgetConv}). 
As reported in Table~\ref{tab:budgetConv}, the SV-PSF IPR algorithm provides more quantitatively correct reconstructions, even with a reduced number of iterations, than classical IPR  and SI-PSF algorithms. Thus, for similar computational time, and if the PSF is calibrated in the FOV, SV-PSF algorithm provides the most quantitative reconstruction of the three methods.

\begin{table}
\centering
\begin{tabular}{|l|c|c|c|c|c|}
\cline{2-6}
\multicolumn{1}{l|}{\makebox[0.1\linewidth]} &
\multicolumn{1}{c|}{{\small{IPR}}} &
\multicolumn{1}{c|}{{\small{SI-PSF IPR}}} &
\multicolumn{3}{c|}{{\small{SV-PSF IPR}}} \\
\hline
$N_{\textmd{iter}}$ & 200 & 200 & 40 & 20 & 10  \\ \hline
$Q$ &  &   & 5 & 10 & 20  \\ \hline 
\small{\# of conv.} & 400 & 400 & 400 & 400 & 400  \\ \hline \hline
\textbf{RMSE} & 0.0755 & 0.0267 & 0.0126 & 0.0114 & 0.0206 \\
\textbf{SSIM}  & 0.1132 & 0.2943 & 0.9853 & 0.9827 & 0.9457  \\
\hline
\end{tabular}
\caption{Reconstruction quality criterion with the 3 hologram formation models with limited convolution budget (400 convolutions). "conv." stands for "convolution".}
\label{tab:budgetConv}
\end{table}

\subsection{Reconstruction results on experimental hologram} \label{sec:ExpeResults}
A similar reconstruction comparison has been performed on experimental hologram. These data have been acquired with the setup described in \cite{brault2022accurate}. The sample consists of calibration latex beads on a microscope slide. The beads are monodispersed of diameter of $1.0 \pm 0.06$ µm and refractive index of $1.58$. The reconstruction of the beads are expected to be similar, regardless of their position in the FOV.
Space-variant aberrations have been added to the hologram recording by choosing a wrong correction collar tuning (0.21 mm instead of 0.17 mm). 
Experimental parameters for the acquisition are provided in Table S1 and pupil functions are displayed in Fig.S3. 
The holograms have been reconstructed on the 204$\times$273 µm$^2$ FOV to highlight the effect of space-varying PSFs. Indeed, for the selected large FOV, both aberrations and defocus exhibit significant variations (Fig. S3).
\begin{figure}[t]
    \centering
    \includegraphics[width=0.84\linewidth]{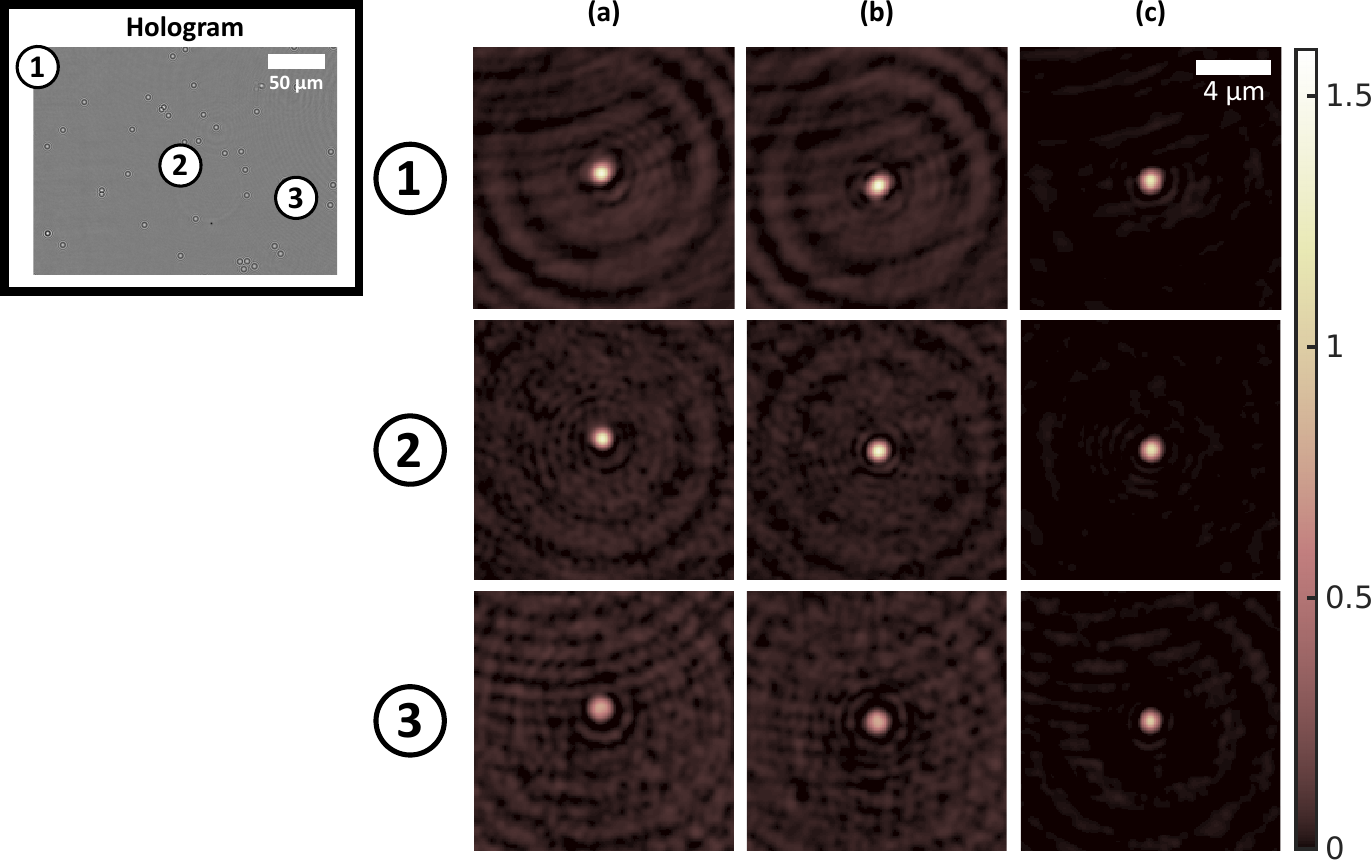}
    \caption{Phase of the reconstruction on three regions in the FOV obtained (a) with classical IPR algorithm, with IPR algorithm (b) with SI-PSF, and (c) with SV-PSF ($P=28,Q=15$)}
    \label{fig:expeRecons}
\end{figure}
Fig. \ref{fig:expeRecons} shows the reconstruction results for the three described methods. The classical IPR algorithm (Fig. \ref{fig:expeRecons}.(a)) poorly succeeds in removing twin-image artifacts for such an aberrated system. The radial symmetry of the reconstructed distribution, expected for these spherical objects, is broken. As seen in region 3, as the PSF varies, the morphological and refractive properties of the object are modified. This can lead to misinterpretation of the reconstruction in other imaging contexts like microbiological ones. 
Using a SI-PSF hypothesis (Fig. \ref{fig:expeRecons}.(b)) slightly improves the reconstruction results. This can be seen in region 2, which is almost at the center of the FOV, where the SI-PSF has been calibrated. Indeed, this reconstruction recovers the expected radial symmetry of the beads. However, this method fails to reconstruct the beads reproducibly when they are located far from the center of the FOV. It should be noted that strong twin image artifacts are still visible on these images as because the selected SI-PSF hypothesis does not compensate for all the space-variant aberrations. 
Using an SV-PSF hypothesis (Fig. \ref{fig:expeRecons}.(c)) highly improves the quality and reproducibility of the reconstruction, as was already demonstrated in the simulations. Most of the twin-image artifacts are removed from the reconstructions and the beads reconstructions are repeatable regardless of position in the FOV.
It should be noted that the reconstruction quality obtained with the SV-PSF models depends on the PSF grid sampling ($P$), the number of modes ($Q$) used to model it, and the number of iterations of the IPR method. Indeed, $P$ and $Q$ should be chosen sufficiently high to capture the variation of the SV-PSF, in our case $P=28, Q=15$ and $N_{iter}=200$. \textcolor{black}{RMSE and SSIM values depending on $Q$ are provided in Fig.S5.} Full FOV reconstructions are provided in Fig.~S4.
\section{Conclusion}
In this paper, we propose a simple way to consider aberration and space-variant PSF in the commonly used IPR algorithm reconstruction. 
Using a calibrated PSF on a regular grid, the propagation model of the IPR algorithm is modified to account for SI-PSF or SV-PSF. When PSFs slowly vary in the FOV, modifying the propagation to account for SI-PSF is sufficient to obtain qualitatively better results than obtained with the classical IPR algorithm with a similar computational cost. However, the SV-PSF should be considered to perform more quantitative reconstructions and reduce the twin image artifacts. This approach relies on a modal decomposition of the SV-PSF. As the image formation model is more accurate, both RMSE and SSIM criteria show better values on simulated and experimental holograms. 
The computational time of the SV-PSF IPR algorithm iterations depends linearly on the number of modes used to capture the variation of the PSF. We have shown on simulated data, that this method leads to more reliable results than the classical IPR algorithm, even with fewer iterations. 
Both methods have been implemented with MATLAB and the source code is provided (see Sec. \ref{sec:code} and Supplementary File).
The proposed aberration correction can be adapted for any other iterative reconstruction algorithms such as the Gerchberg-Saxton algorithm. 
\section{Backmatter}

\textbf{Funding} 
The authors acknowledge for the grant ANR-23-CE51-0023 (project ATICS), and Swiss National Science Foundation research grant 200021\_197107.

\textbf{Disclosures} The authors declare no conflicts of interest.

\textbf{Data Availability Statement}\label{sec:code} The source code and calibration data are available at \url{https://github.com/braultd/Iterative-phase-retrieval-algorithm-for-shift-variant-PSF-in-optical-system-with-aberrations}.

\textbf{Supplemental document}
See Supplementary File for supporting content.


\bibliographystyle{unsrt}
\bibliography{sample}  
\clearpage


\section*{Supplementary material}

# Iterative phase retrieval algorithm for space-variant PSF in optical systems with aberrations - Supplementary file

Dylan Brault[1,*], Corinne Fournier[1], and Tatiana Latychevskaia[2,3]

[1]Université Jean Monnet Saint-Etienne, CNRS, Institut d'Optique Graduate School, Laboratoire Hubert Curien UMR 5516, 42023, Saint-Etienne, France.
[2]Center for Life Sciences, Paul Scherrer Institute, Forschungsstrasse 111, 5232 Villigen, Switzerland.
[3]Department of Physics, University of Zurich, Winterthurerstrasse 190, 8057 Zurich, Switzerland.

*dylan.brault@telecom-st-etienne.fr

Compiled November 25, 2024

---

## SUPPLEMENTARY INFORMATION

### A. Supplementary tables

| | |
|---:|:---|
| Wavelength | 637.6 nm |
| Magnification | 66.5 |
| Numerical Aperture (NA) | 1.2 |
| Pixel pitch | 82.5 nm |
| Total field of view | 204×273 µm² |
| Typical defocus | 17 µm |
| Beads diameter* | (1.0±0.06) µm |
| Beads refractive index* | 1.587 (polystyrene) |
| Refractive index of immersion medium | 1.47 (glycerol) |
| Coverslip thickness | (0.170±0.001) mm |

**Table S1.** Experimental parameters (*from the manufacturer, ThermoFisher Scientific, Inc.)

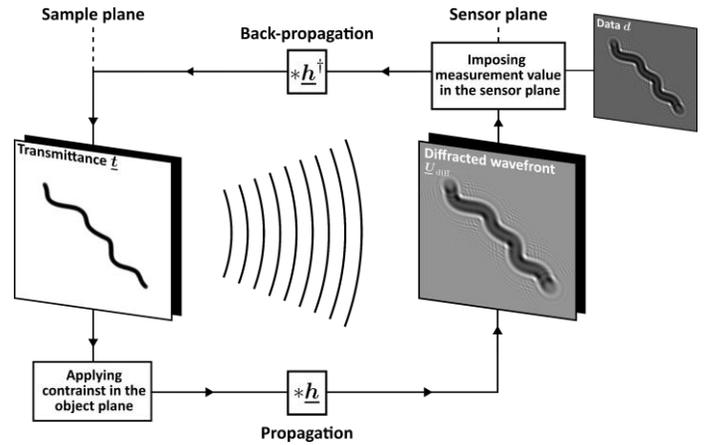

**Fig. S1.** Principle of the IPR phase retrieval algorithm

| Zernike coefficients | $\alpha_2^2$ | $\alpha_3^{-3}$ | $\alpha_3^{-1}$ | $\alpha_3^1$ | $\alpha_3^3$ | $\alpha_4^0$ | $\alpha_4^4$ | $\alpha_5^{-1}$ | $\alpha_6^{-2}$ | $\alpha_6^0$ | $\alpha_6^4$ |
|---|---|---|---|---|---|---|---|---|---|---|---|
| Value for SI-PSF model | 0.62 | −0.25 | 4.91 | −1.28 | 0.19 | −4.71 | 0.45 | −0.71 | 0.15 | −0.8 | 0.20 |

**Table S2.** Zernike coefficients used for the simulations of SI-PSF (all other coefficients are inferior to 0.1)

### B. Supplementary figures

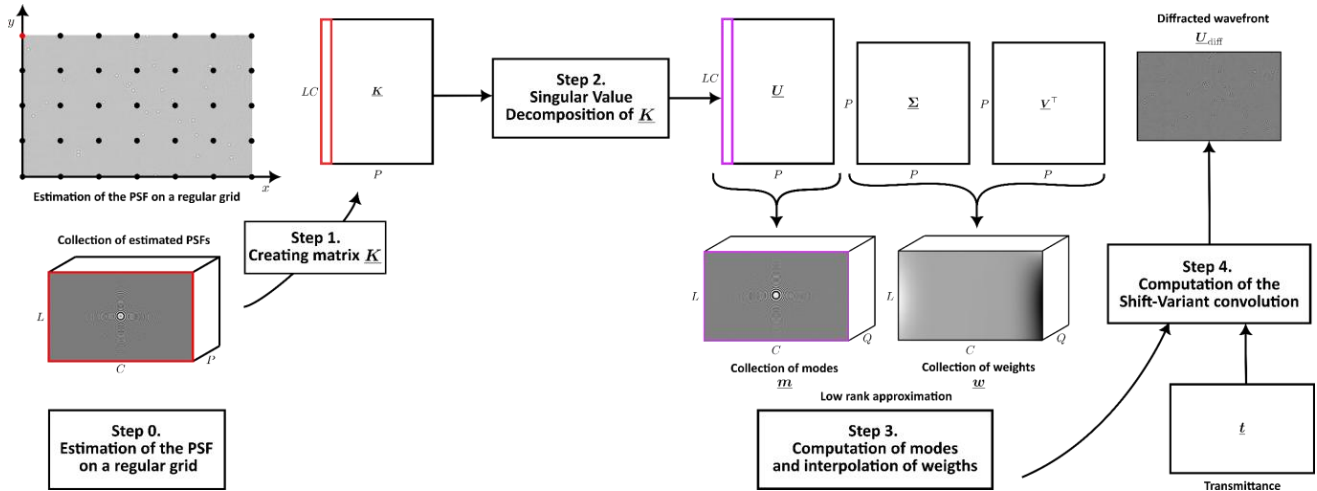

**Fig. S2.** Principle of the PSF interpolation model proposed in [11]

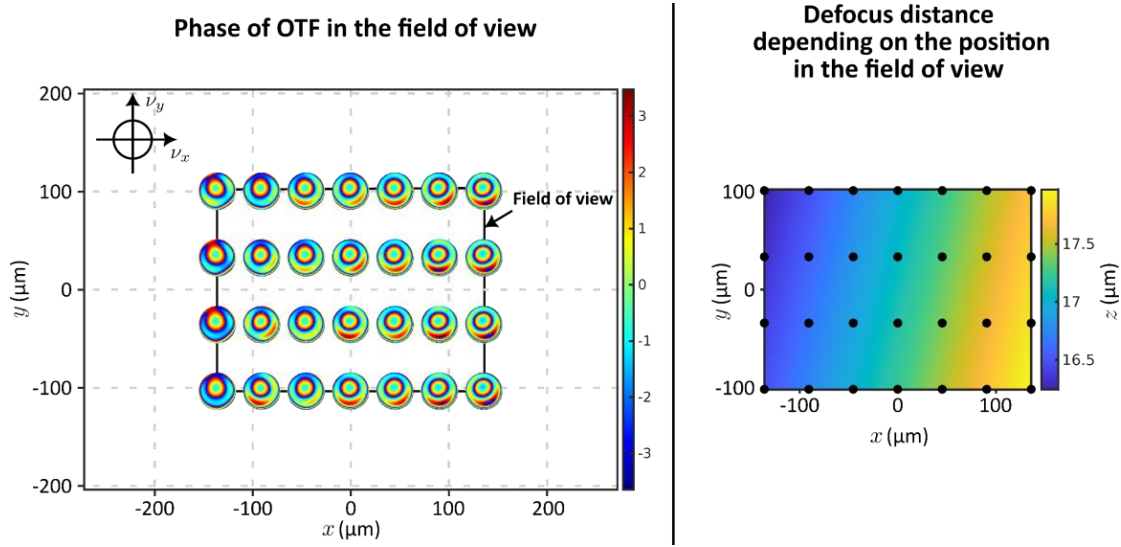

**Fig. S3. Left:** Estimation of the phase of the Optical Transfer Function (OTF) for $z = 0$ depending on the position in the field of view on experimental data, $\nu_x$ and $\nu_y$ are the spatial frequencies in Fourier space. **Right:** Estimation of the defocus depending on the position in the field of view on experimental data. Black dots represent the regular grid on which the OTF and defocus have been estimated. The SV-PSF is computed as a shift-variant convolution between aberrated PSFs in focus and propagation kernels that correspond to the distances z (illustrated on the right side).

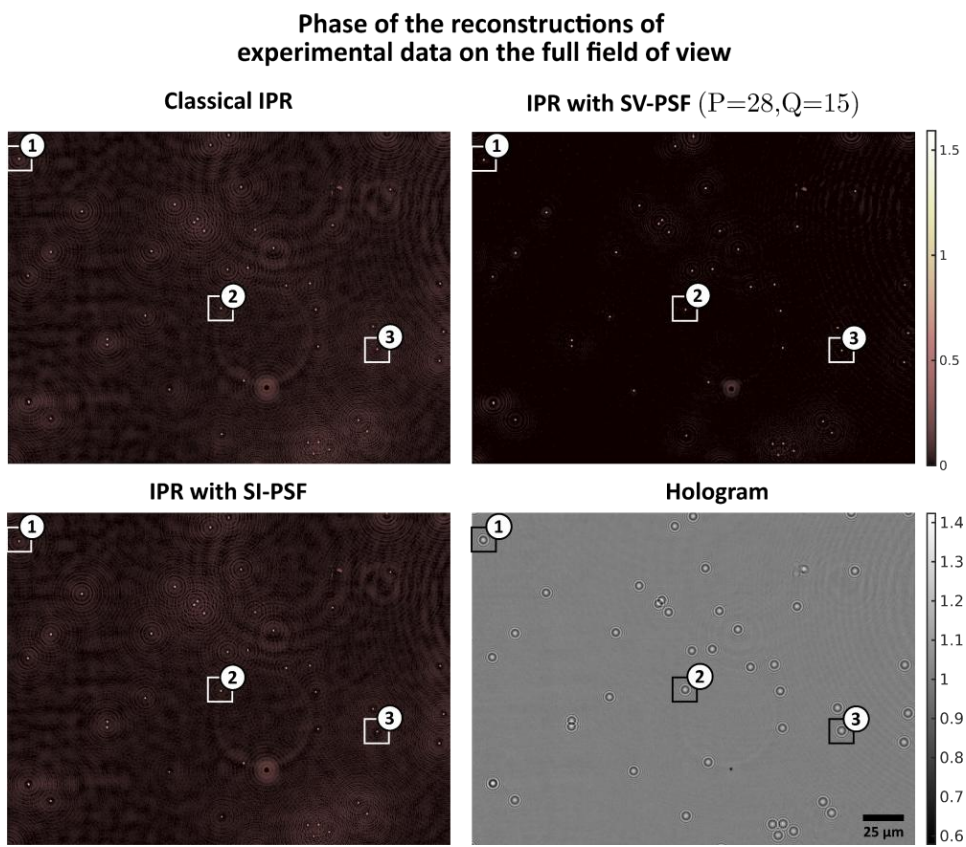

**Fig. S4.** Full field of view reconstructions of an experimental hologram with classical IPR algorithm, IPR algorithm with SI-PSF and IPR algorithm with SV-PSF. Phase of the reconstructions. Zooms on the reconstructions are provided in Fig.2.

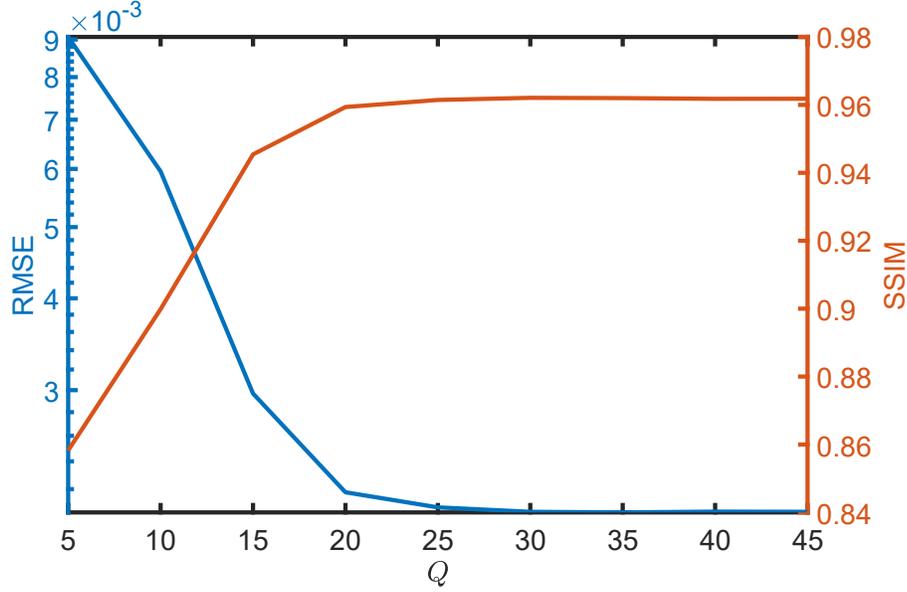

**Fig. S5.** Evolution of the error depending of the $Q$ on simulated data. ($P = 49$)

## C. Implementation remarks

Some implementation remarks are provided in this section to facilitate the implementation of the SV-PSF IPR algorithm.

**Remark #1:** To ensure that SVD does not introduce frequencies beyond $\frac{NA}{\lambda}$ (where $NA$ is the numerical aperture of the microscope objective), each mode $\underline{m}_p$ is filtered in the Fourier domain such that frequencies beyond $\frac{NA}{\lambda}$ are set to 0.

**Remark #2:** As in holography PSFs are spatially extended, padding of the field of view is required to perform convolution. Since holography is not sensitive to phase offset (as only intensity is recorded on the sensor), it is common to choose the phase origin in the object plane such that one-padding can be performed. Thus, the phase of the (0,0) frequency of the propagator is set to 0. In this case:

$$\underline{U}_{\text{diff}} = \underline{t} * \underline{h} = (1 + \underline{o}) * \underline{h} = 1 + \underline{o} * \underline{h} \quad (1)$$

where $\underline{o}$ is a complex-valued function that describes the perturbation caused by the sample. For SV-PSF propagation, the weights $\underline{w}_p$ have to be defined over the padded field of view. As it may be challenging to extrapolate them realistically, we suggest setting them to 0 out of the field of view and propagating only the perturbation $\underline{o}$ such that the pixel-wise multiplication between the weights and $\underline{t} - 1$ or $\underline{U}_{\text{diff}} - 1$ is expected to be 0 outside of the field of view. Finally, as the phase of the (0,0) frequency of the propagator should be set to 0, we set the phase of the (0,0) frequency of each mode $\underline{m}_p$ to 0 for the SV-PSF IPR algorithm.

**Remark #3:**
The modes $\underline{m}_p$ are not the PSF at $p$-th spatial location but an orthogonal set of space-invariant PSF modes, and the $\underline{w}_p$ are the associated weights maps that specify how the modes $\underline{m}_p$ should be weighted together over the FOV. Indeed, the matrix $\underline{K}$ is composed of all the locally calibrated $P$ PSF. As the modes are sorted according to their associated singular value, the low-rank approximation reduces the number of modes required. Indeed, the truncated SVD provides the minimal mean square error between $\underline{K}$ and the approximated matrix of rank $Q$ (Eckart–Young–Mirsky theorem).

## D. Supplementary algorithm

Algorithm 1. Classical IPR algorithm

1: **Input:** Data $\underline{d}$     ($L \times C$ image)
2: **Input:** Propagation kernel $\underline{h}$     ($L \times C$ image)
3: **Input:** Number of iterations $N_{\text{iter}}$     (integer)
4: **Output:** Transmission function $\underline{t}$     ($L \times C$ image)
5:
6: $\underline{U}_{\text{diff}} \leftarrow \sqrt{\underline{d}}$     ↪ *Initialization of the wavefront on the sensor plane*
7:
8: **for** $n_{\text{iter}} = 1$ **to** $N_{\text{iter}}$ **do**
9:     $\underline{t} \leftarrow \underline{U}_{\text{diff}} * \underline{h}^{\dagger}$     ↪ *Back-propagation to the sample plane*
10:
11:     **for** $k = 1$ **to** $L \times C$ **do**
12:       **if** $\underline{t}_k \notin \gamma$ **then**     ↪ *Applying constraints*
13:         $\underline{t}_k = 1$
14:
15:     $\underline{U}_{\text{diff}} \leftarrow \underline{t} * \underline{h}$     ↪ *Propagation to the sensor plane*
16:
17:     $\underline{U}_{\text{diff}} \leftarrow \underline{d} \odot \exp(i \times \arg(\underline{U}_{\text{diff}}))$     ↪ *Updating modulus*
18:
19: **return** $\underline{t}$

# MATLAB CODE

## A. SI-PSF Reconstruction

```
% a_SIPSF_Reconstruction.m
% File to reconstruct a hologram with a SIPSF hypothesis
% The PSF has been calibrated at the center of the field of view

clear all; close all;clc;
% Adding auxiliary functions to path
addpath(genpath('./Utils'))

% Load the hologram load('./Data/data.mat');

% Experimental parameters
pix.nb_x=1000; pix.nb_y=1000;
pix.dx=82.5e-9;
pix.dy=82.5e-9;
lambda=637.6e-9;
n_0=1.47;
ON=1.2;

% Pixel size of the padded images
pix_pad=pix;
pix_pad.nb_x=2*pix.nb_x;
pix_pad.nb_y=2*pix.nb_y;

% Defining functions of padding and cropping
cropimg=@(x) x(pix.nb_y/2+1:pix.nb_y/2+pix.nb_y,pix.nb_x/2+1:pix.nb_x/2+pix.nb_x);
padimg=@(x) padarray(x,[pix.nb_y/2 pix.nb_x/2],1);

% Load the shift-invariant PSF h_SI estimated at the center of the field of view
disp('Loading the shift-invariant PSF...')
load('./Data/PSF_FOV_center.mat')

% Number of iterations of IPR algorithm
n_iter=200;

% Initialization of the wavefront in the hologram plane
disp('Initializing the reconstructed diffracted wavefront...')
U_diff=padimg(sqrt(data));

for n=1:n_iter
    disp(strcat('Processing: ',' ',num2str(floor(100*n/n_iter)),' % ...'))

    % Backpropagation of the wavefront into the sample plane
    t=backpropagation(h_SI,U_diff);


    % Imposing constraints in the sample plane
    mod_t=abs(t);
    arg_t=angle(t);
    arg_t(arg_t<0)=0; % The phase of the sample is positive

    mod_t(:)=1; % The objects are transparent

    t=mod_t.*exp(1i*arg_t);

    % Propagation of the wavefront into the hologram plane
    U_diff=propagation(h_SI,t);
    % Imposing data values in the hologram plane
    mod_U_diff=padimg(sqrt(data));
    arg_U_diff=angle(U_diff);
    U_diff=mod_U_diff.*exp(1i*arg_U_diff);
end
```

```
disp('Recontruction process completed')

% Cropping the extended reconstruction
t=cropimg(t);

% Display the recontructed transmission function
figure(1);
subplot(1,2,1);imshow(abs(t),[]);colormap(pink);colorbar;title('Modulus of the reconstructed
transmittance') subplot(1,2,2);imshow(angle(t),[]);colormap(pink);colorbar;title('Phase of the
reconstructed transmittance')
```

**B. SV-PSF Reconstruction**

```
% b_SVPSF_Reconstruction.m
% File to reconstruct a hologram with a SVPSF hypothesis
% The SVPSF has been calibrated on regular grid

clear all; close all;clc;
% Adding auxialary functions to path
addpath(genpath('./Utils'))

% Load the hologram
load('./Data/data.mat');

% Experimental parameters
pix.nb_x=1000;
pix.nb_y=1000;
pix.dx=82.5e-9;
pix.dy=82.5e-9;
lambda=637.6e-9;
n_0=1.47;
ON=1.2;

% Pixel size of the padded images
pix_pad=pix;
pix_pad.nb_x=2*pix.nb_x;
pix_pad.nb_y=2*pix.nb_y;

% Defining functions of padding and cropping
cropimg=@(x) x(pix.nb_y/2+1:pix.nb_y/2+pix.nb_y,pix.nb_x/2+1:pix.nb_x/2+pix.nb_x);
padimg=@(x) padarray(x,[pix.nb_y/2 pix.nb_x/2],1);

% Load PSFs on regular grid of coordinates X_sub, Y_sub (Each column is a PSF)
disp('Loading the calibrated SV-PSF...')
load('./Data/SV_calibrated_PSFs.mat');

% Coordinated of each pixels of the padded image
[X_field,Y_field]=meshgrid(-pix.nb_x:pix.nb_x-1,-pix.nb_y:pix.nb_y-1);

% Low rank approximation of the PSF
disp('Computing the low-rank approximation of the SV-PSF...')
Rank=15; % Rank of the approximation
[U,S,V]=SVPSF_LowRank(K,Rank);
[m,w]=SVPSF_ConvolutionKernel_WeightsComputation(U,S,V,X_sub,Y_sub,X_field,Y_field);
clear U;clear V;clear S; % Clearing the SVD variable to save memory

% Number of iterations of the IPR algorithm
n_iter=200;
% Initialization of the wavefront in the hologram plane
disp('Initializing the reconstructed diffracted wavefront...')
U_diff=padimg(sqrt(data));
```

```matlab
% Shift-variant propagation and backpropagtion functions
propagation_= @(x) SVPSF_Propagation(m,w,mtfON,x);
backpropagation_=@(x)SVPSF_BackPropagation(m,w,mtfON,x);

for n=1:n_iter
    disp(strcat('Processing: ',' ',num2str(floor(100*n/n_iter)),' % ...'))

    % Backpropagation of the wavefront into the sample plane
    t=backpropagation_(U_diff);

    % Imposing constraints in the sample plane
    mod_t=abs(t);
    arg_t=angle(t);
    arg_t(arg_t<0)=0; % The phase of the sample is positive
    mod_t(:)=1; % The objects are transparent
    t=mod_t.*exp(1i*arg_t);

    % Propagation of the wavefront into the hologram plane
    U_diff=propagation_(t);

    % Imposing data values in the hologram plane
    mod_U_diff=padimg(sqrt(data));
    arg_U_diff=angle(U_diff);
    U_diff=mod_U_diff.*exp(1i*arg_U_diff);

end

disp('Reconstruction process completed')

% Cropping the extended reconstruction
t=cropimg(t);

% Display the recontructed transmission function
figure(1);
subplot(1,2,1);imshow(abs(t),[]);colormap(pink);colorbar;title('Modulus of the reconstructed transmittance') subplot(1,2,2);imshow(angle(t),[]);colormap(pink);colorbar;title('Phase of the reconstructed transmittance')
```

### C. Auxiliary functions

```matlab
function[Uprop]=propagation(propagation_kernel,x)
% This function compute the backpropagation of x using the propagation kernel propagation_kernel
% Inputs:
% propagation_kernel: Propagation kernel
% x: Wavefront to be propagated
% Outputs:
% Uprop: Backpropagated wavefront

    Uprop=ifft2(fft2(propagation_kernel).*fft2(x));
end
```

```matlab
function [Ubprop]=backpropagation(propagation_kernel,x)
% This function compute the backpropagation of x using the propagation kernel propagation_kernel
% Inputs:
% propagation_kernel: Propagation kernel
% x: Wavefront to be backpropagated
% Outputs:
% Ubprop: Backpropagated wavefront

    Ubprop=ifft2(conj(fft2(propagation_kernel)).*fft2(x));
end
```

```matlab
function [ConvolutionKernel,Weights] = 
SVPSF_ConvolutionKernel_WeightsComputation(U,S,V,X_sub,Y_sub,X_field,Y_field)

% This function compute the mode and weights for SV-PSF
% Inputs :
% USV: Singular Value Decomposition of psfs
% X_sub, Y_sub: Coordinates of position of calibrated PSFs (coarse regular grid)
% X_field,Y_field: Coordinates of the pixel grid
% Outputs :
% ConvolutionKernel: Convolution modes for SV convolution
% Weights: Corresponding weights for SV convolution

Weights=[];
ConvolutionKernel=[];

Weights_vec=S*V; % Each column contains weights for one mode on a coarse regular grid
for i=1:size(Weights_vec,1)
    
    % Reshaping the weights on the regular grid
    Weights_sub=reshape(Weights_vec(i,:),[size(X_sub,1),size(X_sub,2)]);
    
    % Interpolation of the weights on the pixel grid
    Weights(:,:,i)=interp2(X_sub,Y_sub,Weights_sub,X_field,Y_field,'cubic',0);
    
    % Reshaping the mode on the pixel grid
    ConvolutionKernel(:,:,i)=reshape(U(:,i),[size(X_field,1),size(X_field,2)]);
    % Each column of U contains a mode
end

end
```

---

```matlab
function [U S V] = SVPSF_LowRank(psfs,rank)
% This function compute the low-rank approximation of the SV-PSF
% Inputs :
% psfs : Calibrated PSFs (each column is a PSF - corresponds to matrix K)
% rank : Rank of the low rank approximation

% Outputs :
% USV: Singular Value Decomposition of psfs (Low Rank Approximation of rank rank)

[U S V]=svd(psfs,'econ');
V_=V';

U=U(:,1:rank);
S=S(1:rank,1:rank);
V=V_(1:rank,:);
end
```

---

```matlab
function [Uprop] = SVPSF_Propagation(ConvolutionKernel,Weights,mtfON,x)

% This function compute the shift-varying propagation
% Inputs :
% ConvolutionKernel : Convolution kernel obtained by SVD decomposition
% Weights : Weigths map associated to the convolution kernel (obtained by SVD decomposition)
% mtfON : MTF correspond to filtering by numerical aperture
% x: Wavefront to be propagated (centered on 1)
% Outputs :
% Uprop: Propagated wavefront
% Supression of the DC component of the signal
x=x-1;
```

```matlab
% Initialization of the wavefront
Uprop=zeros(size(x));

% Low rank approximation of the convolution
for k=1:size(ConvolutionKernel,3)

    % Computation of the Fourier transform of k-th mode
    FTConvKern=fft2(ConvolutionKernel(:,:,k));
    % Setting the phase at the origin of FTConvKern (thus, the phase origin is on the hologram plane)
    FTConvKern(1,1)=abs(FTConvKern(1,1));

    % Computation of the shift-variant convolution
    Uprop=Uprop+ifft2(fft2(Weights(:,:,k).*x).*FTConvKern.*mtfON); end

% Adding DC component of the signal
Uprop=Uprop+1;
end
```

---

```matlab
function [Uprop] = SVPSF_BackPropagation(ConvolutionKernel,Weights,mtfON,x)

% This function compute the shift-varying propagation
% Inputs:
% ConvolutionKernel: Convolution kernel obtained by SVD decomposition
% Weights: Weigths map associated to the convolution kernel (obtained by SVD decomposition)
% mtfON: MTF corresponds to filtering by numerical aperture
% x: Wavefront to be propagated (centered on 1)
% Outputs:
% Uprop: Propagated wavefront

% Suppression of the DC component of the signal
x=x-1;

% Initialization of the wavefront
Uprop=zeros(size(x));

% Low rank approximation of the convolution
for k=1:size(ConvolutionKernel,3)

    % Computation of the Fourier transform of k-th mode (conjugation for back-propagation)
    FTConvKern=conj(fft2(ConvolutionKernel(:,:,k)));
    % Setting the phase at the origin of FTConvKern (thus, the phase origin is on the sample plane)
    FTConvKern(1,1)=abs(FTConvKern(1,1));

    % Computation of the shift-variant convolution (the weights are conjugated for the backpropagation)
    Uprop=Uprop+ifft2(fft2(conj(Weights(:,:,k)).*x).*FTConvKern.*mtfON);
end

% Adding DC component of the signal
Uprop=Uprop+1;
end
```

---


\end{document}